%% file: dis.tex
\documentstyle[12pt, UCLAPhD]{report}

\begin{document}

%
%
\input{macros.tex}

%
%
\input{prelim.tex}

%
%
\chapter{Introduction}
\label{cintro}
\input{intro.tex}

%
%
\chapter{Seiberg-Witten Theory}
\input{sw.tex}

%
%
\chapter{SW theory with fundamental hypermultplets}
\input{fund.tex}

%
%
\chapter{SW theory with an adjoint hypermultiplet}
\input{adj.tex}

%
%
\chapter{Conclusion}
\input{conclusion.tex}

%
%
\input{bib.tex}

%
%
\end{document}

%% file: macros.tex
%
%
\def\F{{\cal F}}
\def\A{{\cal A}}
\def\N{{\cal N}}
\def\D{\partial}
\def\L{\bar{\Lambda}}
\def\sL{\Lambda}
\def\ab{\bar a}
\def\akb{\bar a_k}
\def\alb{\bar a_l}
\def\amb{\bar a_m}
\def\anb{\bar a_n}
\def\bk2j{\beta_k^{(j)}}
\def\bktwo{\beta_k^{(1)}(a)}
\def\bkfour{\beta_k^{(2)}(a)}
\def\bksix{\beta_k^{(3)}(a)}
\def\bkeight{\beta_k^{(4)}(a)}
\def\bkten{\beta_k^{(5)}(a)}
\def\bktwelve{\beta_k^{(6)}(a)}
\def\bkfourteen{\beta_k^{(7)}(a)}
\def\bksixteen{\beta_k^{(8)}(a)}
\def\btwo{\beta^{(1)}}
\def\bfour{\beta^{(2)}}
\def\bsix{\beta^{(3)}}
\def\beight{\beta^{(4)}}
\def\Dkj{\Delta_k^{(j)}}
\def\Dtwo{\Delta^{(1)}}
\def\Dfour{\Delta^{(2)}}
\def\Dsix{\Delta^{(3)}}
\def\rDzero{\Delta^{(0)}}
\def\rDtwo{\Delta^{(1)}}
\def\rDfour{\Delta^{(2)}}
\def\rDsix{\Delta^{(3)}}
\def\rDeight{\Delta^{(4)}}
\def\rDten{\Delta^{(5)}}
\def\rDtwelve{\Delta^{(6)}}
\def\rDfourteen{\Delta^{(7)}}
\def\rDsixteen{\Delta^{(8)}}
\def\onesum{\sum_{k=1}^{r}}
\def\twosum{\sum_{k,m=1}^{r}}
\def\threesum{\sum_{k,m,n=1}^{r}}
\def\foursum{\sum_{k,m,n,l=1}^{r}}
\def\fivesum{\sum_{k,m,n,l,o=1}^{r}}
\def\oneden{\D a_k}
\def\twoden{\D a_k \D a_m }
\def\threeden{\D a_k \D a_m \D a_n}
\def\fourden{\D a_k \D a_m \D a_n \D a_l}
\def\fiveden{\D a_k \D a_m \D a_n \D a_l \D a_o}
\def\instone{\F^{(1)} }
\def\insttwo{\F^{(2)} }
\def\instthree{\F^{(3)} }
\def\instfour{\F^{(4)} }
\def\instfive{\F^{(5)} }
\def\instsix{\F^{(6)} }
\def\instseven{\F^{(7)} }
\def\insteight{\F^{(8)} }

%% file: prelim.tex
%
%

\Title{Nonperturbative Recursion Relations in $\N=2$ Supersymmetric Gauge Theories}
\Author{Gordon James Chan}
\DegreeDate{2001}
\Major{Physics}

%
%

\CommitteeMember{Zvi Bern}
\CommitteeMember{David Gieseker}
\CommitteeMember{E. Terry Tomboulis}
\CommitteeChair{Eric D'Hoker}

%
%

\Dedication
{
\centering{\bf This Thesis Is Dedicated To}
\\
\centering{\bf Jon C.}
\\
}

%
%

\Acknowledgements
{ 
I am very grateful to my advisor Eric D'Hoker for his guidance and 
wisdom when it comes to physics and various life issues, and to 
Zvi Bern, Christian Fronsdal, David Gieseker, and E. Terry Tomboulis
for being on my committee.
I benefited greatly from my acquaintance and friendships 
with (in no particular order)
Amir Ahsan,
Nina Byers,
Elena Caceres,
Gordon Chalmers,
Sergey Cherkis,
Robert Finkelstein,
Darius Gagne,
Eran Marcus,
Joel Rozowsky,
Anton Ryzhov,
Gabriele Varieschi,
and Kingston Wang.
}

%
%

\Vita
{
\VitaItem{May 28, 1973}{Born in Ottawa, Ontario, Canada}
\VitaItem{1995}
{
Bachelor of Science
\\
in Physics
\\
Carleton University (Ottawa, Canada)
}
\VitaItem{1995-1997}
{
NSERC PGS-A Fellowship (Canada)
}
\VitaItem{1997}
{
Master of Science
\\
in Physics
\\
University of California, Los Angeles
}
\VitaItem{1997-1999}
{
NSERC PGS-B Fellowship (Canada)
}
\VitaItem{}{}
\VitaItem{}{}
\VitaItem{}{}
\VitaItem{}{}
\VitaItem{}{}
\VitaItem{}{}
\VitaItem{}{}
\VitaItem{}{}
\VitaItem{}{}
\VitaItem{}{}
\VitaItem{}{}
\VitaItem{}{}
}

%
%

\Publication
{
Gordon Chan, Robert Finkelstein, and Vadim Oganesyan.
"The q-Isotropic Oscillator".
J. Math. Phys. 38(5), 2132, (1997).
} 
\Publication
{
Gordon Chan, Eric D'Hoker.
"Instanton Recursion Relations for the Effective Prepotential in
${\cal N}=2$ Super Yang-Mills".
Nucl. Phys. B 564, 503, (2000).
}
\Publication
{
Gordon Chan.
``Prepotential Recursion Relations in ${\cal N}=2$ Super Yang-Mills
with Adjoint Matter''.
UCLA/01/TEP/8, hep-th/0105302.
}

%
%

\Abstract{Nonperturbative Recursion Relations in $\N=2$ Supersymmetry
Gauge Theories}
{
Linear recursion relations for the instanton corrections to the
effective prepotential are derived for two cases of $\N=2$ 
supersymmetric gauge theories; the first case with an arbitrary number of 
hypermultiplets in the fundamental representation of an arbitrary
classical gauge group, and the second case with one hypermultiplet 
in the adjoint representation of $SU(N)$.  The
construction for both cases proceed from the Seiberg-Witten solutions
and the renormalization group type equations for the prepotential.
Successive iterations of these recursion relations allow us to simply
obtain instanton corrections to an arbitrarily high order.  Checks
with previous results in the literature were performed.  Other
theoretical properties and generalizations are also discussed.
}

%
%

\MakePreliminaryPages

%% file: intro.tex
\setcounter{section}{0}
\setcounter{subsection}{0}
\setcounter{equation}{0}
\setcounter{footnote}{0}

%

Over the past half decade there has been great progress in understanding
the non-perturbative dynamics of $\N$=2 SUSY gauge theories starting with 
the SU(2) case \cite{SW1}\cite{SW2}, with further generalizations 
to other gauge groups \cite{Yank}-\cite{Argyres} with
additions of matter hypermultiplets \cite{HanOz}-\cite{PKsoliton}.
Non-perturbative corrections in weak coupling corresponding to instanton
effects \cite{Seiberg88} were evaluated using field theory 
techniques to one instanton 
\cite{Seiberg88}\cite{ShiftVain}\cite{Poul}\cite{Trav} 
and two instanton \cite{Dor1}\cite{Dor2} orders.  
Some of the previous instanton calculations using the Seiberg-Witten ansatz
were performed by solving the Picard-Fuchs equations for the 
period integrals corresponding to the quantum moduli parameters representing
the set of vacuum expectation values of the Higgs fields 
\cite{Lerche}\cite{Ohta1}\cite{Ohta2}\cite{Ewen1}\cite{Ewen2}.
Other previous calculations involved solving the period integrals directly
\cite{DPKi}, and were found to be in agreement with the Picard-Fuchs and
field theory results.
In an intriguing paper \cite{Matone}, a recursion relation for the 
instanton corrections to the effective prepotential $\F$ was found for 
the pure $SU(2)$ case which led us to seek a generalization of this result 
for any gauge group and number of matter hypermultiplets.

In a related development, the Seiberg-Witten equations were viewed 
analogously to the Whitham hierarchy equations 
\cite{Whitham1}-\cite{Whitham5}
and the WDVV equations \cite{WDVV1}\cite{WDVV2}.  
Nonlinear recursion relations for
the instanton corrections involving Jacobi $\theta$-functions (which 
themselves involve $\tau_{ij}$ as in \cite{DPKr}) were derived starting
from the Whitham hierarchy equations 
\cite{SpanW1}-\cite{SpanW3}. The beta function of the 
prepotential $\F$, first observed in \cite{Sonnen} and later proved in
\cite{DPKr}, provides a very
direct way at calculating instanton corrections to the prepotential
$\F$ without having to perform complicated hyperelliptic integrals and
immediately obtaining rational expressions without $\theta$-functions.
The Seiberg-Witten solutions for classical gauge groups $SU(N)$,
$SO(N)$, and $Sp(N)$ with matter hypermultiplets in the fundamental
representation of the gauge group and the 
renormalization group like equation for the prepotential $\F$,
led us to the discovery of a general recursion relation expressing
the $n-th$ order instanton correction to the prepotential $\F$
in terms of the $(n-1)th, \cdots , 1st$ order instanton corrections 
\cite{Recur}.

In a further development, 
connections between Seiberg-Witten theory and integrable systems were
discovered first for the case of pure $\N$=2 super Yang-Mills theory
in connection with Toda lattices \cite{MarWar}
and Whitham theory \cite{Whitham1}-\cite{Whitham5}.  Later connections between
$\N$=2 super Yang-Mills theory with one hypermultiplet in the adjoint
representation of the gauge group, and the Hitchin \cite{Donagi}
and Calogero-Moser \cite{Martinec} integrable systems was made.  (There are
claims that the Calogero-Moser integrable system can be derived from 
the Hitchin integrable system \cite{Markman}).  Convenient parameterizations
of the Calogero-Moser integrable system useful for performing explicit
Seiberg-Witten type of calculations were discovered \cite{Calegro}.
The Calogero-Moser construction \cite{Calegro} of the Seiberg-Witten 
solution for $\N$=2 super Yang-Mills theory with
one hypermultiplet in the adjoint representation of the gauge
group $SU(N)$ and the renormalization group like
equation for the prepotential $\F$, led us to the discovery of a general
recursion relation expressing the n-th order instanton correction to the
prepotential $\F$ in terms of the (n-1)-th, ..., first order instanton
corrections \cite{CalegroRecur}.

We start off this thesis by reviewing the original Seiberg-Witten problem  
for $\N$=2 super Yang-Mills theory with an $SU(2)$ gauge group with
no hypermultiplets.  Then we will explore extensions to 
the cases with hypermultiplets in the fundamental representation of 
any arbitrary classical gauge group, along with explicit calculations for 
several special $SU(2)$ and $SU(3)$ cases.  Lastly we will discuss 
the case of an adjoint hypermultiplet with gauge group $SU(N)$,
along with explicit calculations for the $SU(2)$ case and 
discussions of S-duality.  A summary of the recursion relations
methodology in Seiberg-Witten theory and possible generalizations are 
discussed in the conclusion.

%% file: sw.tex
\setcounter{section}{0}
\setcounter{subsection}{0}
\setcounter{equation}{0}

The Seiberg-Witten (SW) ansatz gives a prescription for
determining the prepotential of the effective action for
$\N=2$ supersymmetric Yang-Mills gauge theories, as
well as for determining the spectrum of BPS states 
\cite{SW1}.

\vspace{7mm}
\addtocounter{section}{1}
{\large \bf \thesection.  $\N=2$ Supersymmetric Gauge Theories}
\medskip

$\N=2$ supersymmetric gauge theories are constructed out
of an $\N=2$ chiral multiplet with an optional 
$\N=2$ hypermultiplet.  
In terms $\N=1$ chiral and vector multiplets,
the $\N=2$ chiral multiplet 
consist of one $\N=1$ vector multiplet and one $\N=1$
chiral multiplet all in the adjoint representation of
the gauge group with the following $\N=1$ superfield content:
\begin{eqnarray}
W_\alpha & \rightarrow & (A_\mu, \lambda, D) \\ \nonumber
\Phi & \rightarrow & (\phi, \psi, F) \nonumber
\end{eqnarray}
where $A_\mu$ is a gauge boson, 
$\lambda , \psi$ are  Weyl fermions,  
$\phi$ is a scalar field, and $D , F$ are auxillary fields.
The $\N=2$ hypermultiplet consists of two $\N=1$ chiral multiplets
all in a representations of the gauge group with the following
$\N=1$ superfield content:
\begin{eqnarray}
Q & \rightarrow & (\psi_q, q, F) \\ \nonumber
\tilde{Q} & \rightarrow & 
(\psi_{\tilde{q}}^\dagger, \tilde{q}^\dagger, \tilde{F}  ) 
\nonumber
\end{eqnarray}
where $\psi_q , \psi_{\tilde{q}}^\dagger$ are Weyl fermions,
$q , \tilde{q}^\dagger$ form a complex scalar, and $F , \tilde{F}$ are
auxillary fields. 

The most general renormalizable $\N=1$ action 
consistent with the $\N=2$ supersymmetry field content
in $\N=1$ superfield
notation takes on the form
(in Weinberg's notation \cite{Weinberg3})
\begin{eqnarray}
{\cal L} & = & {\cal L}_{chiral} + {\cal L}_{hyper} \\ \nonumber
{\cal L}_{chiral} & = & \frac{1}{2} \Phi^\dagger e^{-2V} \Phi \mid_D
-\frac{1}{2} [\frac{\tau}{8\pi i} W_\alpha W^\alpha \mid_F + h.c. ]
\\ \nonumber
{\cal L}_{hyper} & = & \sum_{i=1}^{N_f} 
\frac{1}{2} 
\{ 
[Q^\dagger_i e^{-2V} Q_i + \tilde{Q}_i e^{2V} \tilde{Q}^\dagger_i]
\mid_D 
+ 
[\tilde{Q}_i \Phi Q_i + m_i \tilde{Q}_i Q_i] \mid_F + h.c.
\}
\end{eqnarray} 
where the D and F integrations are over $d^4 \theta$ and $d^2 \theta$
respectively, and 
\begin{eqnarray}
\tau & = & \frac{4\pi i}{g^2} + \frac{\theta}{2\pi} \label{eq:complexg}
\end{eqnarray}
is the complexified gauge coupling constant.  In component form, the
bosonic part of the action with the auxillary fields eliminated 
using the Euler-Lagrange equations of motion will take on the form
\begin{eqnarray}
{\cal L} & = & (D_\mu \phi)^\dagger D^\mu \phi 
-\frac{1}{4}F_{\mu \nu} F^{\mu \nu} + 
\frac{g^2 \theta}{32 \pi^2} F_{\mu \nu} \tilde{F}^{\mu \nu}
- V(\phi, \phi^\dagger) \\ \nonumber
& + & (interactions)
\end{eqnarray}
where
\begin{eqnarray}
V(\phi, \phi^\dagger) = \frac{1}{2} tr([\phi^\dagger, \phi])^2
\end{eqnarray}
To minimize the potential $V(\phi, \phi^\dagger)$ without setting all the
expectation values of all the scalar fields to zero, one can perform
a spontaneous symmetry breaking of the gauge symmetry by keeping the
scalar fields in the Cartan subalgebra of the gauge group and giving them
a non-zero expectation value, while all the other scalar fields are given
a zero expectation value.  (For an $SU(N_c)$ gauge group, this breaks
the gauge symmetry down to a $U(1)^{N_c-1}$ Coulomb phase).  This process
gives masses to the fields outside of the Cartan subalgebra and breaks
$\N=2$ supersymmetry.  In order to preserve $\N=2$ supersymmetry, 
one integrates out all the massive fields outside of the Cartan
subalgebra, which will restore $\N=2$ supersymmetry in the remaining
Cartan subalgebra fields but will also add in non-renormalizable 
terms consistent with the other symmetries in the theory.  
Additionally, an anomaly along with this
spontaneous symmetry breaking and subsequent integrating out prescription,
breaks the classical ${\cal R}$-symmetry group of the 
theory to a subgroup of the original classical ${\cal R}$-symmetry \cite{SW1}
\cite{Weinberg3}.  

The most general non-renormalizable $\N=1$ supersymmetric action 
consistent with the $\N=2$ chiral multiplet field content
in $\N=1$ superfield notation takes on the form
\begin{eqnarray}
{\cal L}_{chiral} & = & 
\frac{1}{2}[ K(\Phi, \Phi^\dagger e^{-2V}) ] \mid_D
-\frac{1}{2}[h (\Phi) W_\alpha W_\alpha \mid_F + h.c. ]
\end{eqnarray}
where $h(\Phi)$ is a holomorphic coupling that is equal to
$\frac{\tau}{4\pi i}$ at tree level.
In general, a 1PI effective action of this form 
will have non-holomorphic
contributions to $\tau$ coming from the IR if there are massless
particles in the theory \cite{ShiftVain}.  
One way around this is to take the Wilson effective action
where all massive fields with mass greater than a scale $a$
\footnote{In general, $a$ is taken to be smallest of
the expectation values of the Higgs scalar fields in the Cartan
subalgebra. The masses given to the massless fields from 
spontaneous symmetry breaking are of the order of the expectation
value of the Higgs scalar fields in the Cartan subalgebra.}
are integrated out and
in the loop integrals, the region of integration is over
momenta $\Lambda < | k | < a $ with $\Lambda \ll a$, 
where $\Lambda$ is the scale generated from dimensional transmutation
like an IR cutoff while $a$ acts
like a UV cutoff.  This Wilson effective action preserves the
holomorphicity of the coupling constant $\tau$ \cite{ShiftVain}. 

Imposing $\N=2$ supersymmetry explicitly requires
\begin{eqnarray}
K (\Phi, \Phi^\dagger) = Im [\Phi^\dagger 
\frac{\partial {\cal F} (\Phi)}{\partial \Phi}] & , &
h_{AB}(\Phi) = 
\frac{\partial^2 {\cal F}(\Phi)}{\partial \Phi_A \partial \Phi_B}
\end{eqnarray}
producing a Wilson effective action of the form
\begin{eqnarray}
{\cal L}_{chiral} & = & \frac{1}{8\pi} Im \{
[\frac{\partial {\cal F}(A)}{\partial A^i} \bar{A}^i] \mid_D
+\frac{1}{2}
[\frac{\partial^2 {\cal F}(A)}{\partial A^i \partial A^j}
W_i W_j] \mid_F \} \label{eq:lagrangian}
\end{eqnarray}
where ${\cal F}$ is a holomorphic prepotential that is the
object of interest for which Seiberg-Witten \cite{SW1} provided a
prescription for calculating it.

\vspace{7mm}
\addtocounter{section}{1}
{\large \bf \thesection.  Seiberg-Witten ansatz for pure $SU(2)$ super
Yang-Mills Theory}
\medskip

Previous nonperturbative calculations 
performed in SUSY gauge theories using field theory methods
\cite{Seiberg88}\cite{ShiftVain} were done as weak coupling
expansions around instanton solutions, which contributed as
corrections to the prepotential $\F$ for the $\N=2$ SUSY cases.

For an $\N=2$ super Yang-Mills theory with 
gauge group $SU(2)$ and no
hypermultiplets that's spontaneously broken as prescribed in the
last section, the prepotential in the Wilson effective action
will have the general form
\begin{eqnarray}
\F (a) & = & \frac{i}{2\pi} a^2 ln (\frac{a^2}{\Lambda}) + 
\sum_{j=1}^{\infty} b_m (\frac{\Lambda}{a})^{4m} \label{eq:su2prepot}
\end{eqnarray}
where $a$ is the expectation value of the remaining Higgs field left
that's in the Cartan subalgebra of $SU(2)$, and $\Lambda$ is the scale
produced by the one-loop quantum corrections via renormalization.
The first term in (\ref{eq:su2prepot}) are the contributions from 
the tree level and one-loop terms, while the second sum 
consists of the sum of instanton factors
\begin{eqnarray}
(\frac{\Lambda}{a})^{4m} = e^{m2\pi i \tau}
\end{eqnarray}
weighted by $b_m$'s, due to the $m-th$ instanton.  $\tau$ is the
complexified coupling constant (\ref{eq:complexg}).

Seiberg-Witten in \cite{SW1} show
that the instanton sum in (\ref{eq:su2prepot}) is indeed 
nonvanishing, and uses that fact to derive a prescription
to exactly solve for the $b_m$'s without using field theory methods.
To show this, one starts off by defining 
a Kahler metric on the moduli space
\begin{eqnarray}
ds^2 & = & Im [ \tau da d{\bar a} ] \\ \label{eq:modulispace} \nonumber
& = & Im [ da_D d{\bar a} ] 
\end{eqnarray}
where
\begin{eqnarray}
\tau  =  \frac{\D^2 \F}{\D a^2} & , & 
a_D  =  \frac{\D \F}{\D a} \label{eq:adref}
\end{eqnarray}
Taking the first term in (\ref{eq:su2prepot}) and substituting it
into (\ref{eq:adref}), we get
\begin{eqnarray}
a_D & = & \frac{2ia}{\pi} ln(\frac{a}{\Lambda}) + \frac{ia}{\pi} 
\label{eq:duala}
\end{eqnarray}
which is representative of the high energy perturbative part of 
the asymptotically free theory.  A change of 
variables $u = \frac{1}{2} a^2$ is performed to identify in 
the moduli space the gauge equivalent moduli $a \rightarrow -a$
via the Weyl symmetry of $SU(2)$.  
The moduli space represented by the perturbative part of the theory
is the region $u \rightarrow \infty$, where under a monodromy at
large $u \rightarrow \infty$ produces
\begin{eqnarray}
ln(a) & \rightarrow & ln(a) + \pi i \\ \nonumber
ln(u) & \rightarrow & ln(u) + 2 \pi i
\end{eqnarray}
which induces a change in the moduli 
\begin{eqnarray}
a_D & \rightarrow & -a_D + 2a \label{eq:largeu} \\
a & \rightarrow & -a \nonumber
\end{eqnarray}
using (\ref{eq:duala}).
Representing the moduli as a vector
\begin{eqnarray}
v & = & 
\left(
\begin{array}{c}
a_D \\ a 
\end{array}
\right) \label{eq:mvector}
\end{eqnarray}
the moduli change (\ref{eq:largeu}) has the form
\begin{eqnarray}
v \rightarrow M_\infty v &, & M_\infty = 
\left(
\begin{array}{rr}
-1 & 2 \\
0 & -1 
\end{array}
\right)
\end{eqnarray}
There mere fact that $M_\infty$ is not the identity, implies the
existence of singularities in the moduli space.  If the metric on
the moduli space 
$Im [\tau(a)]$ is globally defined it won't be positive definite because
the harmonic function $Im [\tau(a)]$ can't have a minimum.  Hence the
metric can only be defined locally at these singularities in the moduli
space.

A closer examination of the moduli space metric (\ref{eq:modulispace})
reveals that it's invariant under $v \rightarrow Mv$ transformations, 
if $M \in SL(2, R)$.  In general, $SL(2, R)$ is generated by the 
basis
\begin{eqnarray}
T_b = 
\left(
\begin{array}{rr}
1 & b \\
0 & 1 
\end{array}
\right)
& , &
S =
\left(
\begin{array}{rr}
0 & 1 \\
-1 & 0
\end{array}
\right)
\end{eqnarray}

Under a $T_b$ transformation, the moduli transforms as
\begin{eqnarray}
a_D & \rightarrow & a_D + ba \\ \nonumber
a & \rightarrow & a
\end{eqnarray}
which corresponds to a transformation of
$\theta \rightarrow \theta + 2\pi b$ for the $\theta$ angle in the
Lagrangian (\ref{eq:lagrangian}).  For invariance of 
(\ref{eq:lagrangian}) under $T_b$, we
require $b \in Z$ and hence the moduli transformations are invariant
under $SL(2, Z)$.

Under an $S$ transformation, the moduli transforms as
\begin{eqnarray}
a_D & \rightarrow & a \\ \nonumber
a & \rightarrow & -a_D
\end{eqnarray}
which is realized in the magnetic sector of the theory, where a
magnetic monopole is coupled to a dual magnetic vector field.  The
equivalent Wilson effective action was derived \cite{SW1} to be
\begin{eqnarray}
{\cal L}_{chiral} & = & \frac{1}{8\pi} Im \{
[\frac{\partial {\cal F}_D (A_D)}{\partial A_D} \bar{A_D}] \mid_D
+\frac{1}{2}
[\frac{\partial^2 {\cal F}_D (A_D)}{\partial A_D^2}
W_D^2] \mid_F \} \label{eq:dlagrangian}
\end{eqnarray}
where
\begin{eqnarray}
\frac{\partial^2 {\cal F}_D (A_D)}{\partial A_D^2} & = &
\tau_D \\ \label{eq:dualconst}
& = & -\frac{1}{\tau} \nonumber
\end{eqnarray}
Equation (\ref{eq:dlagrangian}) is like a ``dual'' magnetic 
theory coupled to magnetic monopoles, with a coupling constant
$\tau_D = -\frac{1}{\tau}$ exhibiting the $S$ transformation of
the original Lagrangian (\ref{eq:lagrangian}).  

These $SL(2, Z)$ transformations are in general not a ``symmetry'' but
a way of transforming between different local descriptions of the
moduli space.  A fundamental representation of $SL(2, Z)$ will
be a basis of monodromy matrices including $M_\infty$.
If the monodromy matrices 
representing the singularities commuted with $M_\infty$, then that
would make $a^2$ a good global coordinate with a global harmonic 
function $Im [\tau(a)]$ representing the moduli space.  But previously 
it was argued that this metric will not be positive definite.  Hence
the other option is that the monodromy matrices do not commute 
with $M_\infty$, and will form a non-abelian representation for
$SL(2, Z)$.

In the original $SU(2)$ $\N=2$ super Yang-Mills theory,
there's a classical $U(1)_{\cal R}$ ${\cal R}$-symmetry 
that gets broken down to $Z_8$ by an anomaly.  Under the spontaneous
symmetry breaking and integrating out prescription outlined in the
previous section, the $Z_8$ gets further broken down to $Z_4$.  Under
the change of coordinates $u = \frac{1}{2}a^2$, there is a
$Z_2$ ${\cal R}$-symmetry remaining.  In a minimal non-abelian representation
of $SL(2, Z)$, two singularities in the 
$u$-plane of the moduli space ``related'' to one another by 
$u \rightarrow -u$ along with the singularity at infinity are conjectured
to produce a fundamental representation of for $SL(2, Z)$.
Seiberg-Witten proposed \cite{SW1} that the two 
strong coupling singularities in the
$u$-plane of the moduli space are that of the massless charged 
particle BPS states.  

In $SU(2)$ supersymmetric Yang-Mills theory, in general there are electrically
and/or magnetically charged particles of mass $M$ and charge $Z$
defined as
\begin{eqnarray}
Z & = & n_m a_D + n_e a \\
M & \geq & \sqrt{2} |Z| 
\end{eqnarray}
where $n_m$ and $n_e$ are the number of magnetic and electric charges
respectively.  The BPS states saturate the bound 
$M = \sqrt{2} |Z|$.

For the massless magnetic monopole, it will have
charges $(n_m, n_e) = (1,0)$ in a region around $a_D = 0$ in the
moduli space with a mass $M=0$.  Taking $u=1$ as the point in the
moduli space for the massless monopole, the surrounding region around
it can be parameterized as $a_D \approx c(u-1)$ where $c$ is a constant.
In the dual magnetic sector of the theory (\ref{eq:dlagrangian}), the
magnetic coupling constant at tree level will be
\begin{eqnarray}
\tau_D & = & \frac{\D^2 \F_D (a_D)}{\D a_D^2} \\ 
& \approx & -\frac{i}{\pi} ln(a_D) \nonumber
\end{eqnarray}
which integrates to
\begin{eqnarray}
a & = & \frac{\D \F_D (a_D)}{\D a_D} \\ 
& \approx & a_0 + \frac{i}{\pi} a_D ln(a_D) \nonumber \\
& \approx & a_0 + \frac{i}{\pi} c(u-1) ln[c(u-1)] \nonumber
\end{eqnarray}
where $a_0$ is a constant.
A monodromy around $u=1$ will transform as
\begin{eqnarray}
ln[c(u-1)] & \rightarrow & ln[c(u-1)] + 2\pi i \\ 
a_D & \rightarrow & a_D \\ \nonumber
a & \rightarrow & a - 2 a_D \nonumber
\end{eqnarray}
where in terms of the moduli space vector $v$ (\ref{eq:mvector})
\begin{eqnarray}
v \rightarrow M_1 v & , & M_1 =
\left(
\begin{array}{rr}
1 & 0 \\
-2 & 1 
\end{array}
\right)
\end{eqnarray}
If we take the charges of particle as a row vector
\begin{equation}
q = (n_m, n_e) \label{eq:qvector}
\end{equation}
for which the massless magnetic monopole will
have $q_1 = (1,0)$, the monodromy matrix $M_1$ will have
an invariant transformation $q_1 M_1 = q_1$ with the charge vector $q_1$.

To examine what the third singularity is, the monodromy at 
$u \rightarrow \infty$ can be decomposed into the monodromies around
the $u=1$ and the third singularity denoted as $u=-1$.  In terms of
the monodromy matrices, it will take on the form 
$M_\infty = M_1 M_{-1}$ which produces
\begin{eqnarray}
M_{-1} & = &
\left(
\begin{array}{rr}
-1 & 2 \\
-2 & 3 
\end{array}
\right) \label{eq:3sing}
\end{eqnarray}
In terms of charges represented by (\ref{eq:qvector}), we get a
massless dyon
$q_{-1} = (1, -1)$ being invariant under $q_{-1}M_{-1} = q_{-1}$.

Hence the monodromy matrices $M_\infty, M_1, M_{-1}$ form a
subgroup $\Gamma(2)$ of $SL(2, Z)$.  The solution to this math
problem is in \cite{Clemens} and used by Seiberg-Witten \cite{SW1} as
\begin{eqnarray}
a  =  \oint_A d\lambda & , &
a_D = \oint_B d\lambda \nonumber \\
d\lambda = \frac{\sqrt{2}}{2\pi} \frac{(x-u)dx}{y} & , &
y^2 = (x-1)(x+1)(x-u)
\end{eqnarray}
where the contour integrals are over a Riemann surface with
branch cuts between -1 and 1 on the real x-axis and between
$u$ to $\infty$ in the complex x-plane.  The integral
for $a$ has a closed contour A taken around the branch cut between
x=-1 and x=1, while the integral for $a_D$ has a closed contour B taken
through the branch cuts at x=1 and x=u.  (The topology of this
Riemann surface is a torus with the A and B contours taken as
the homology cycles, where the B contour is around the center hole
of the torus \cite{Clemens}). These integrals
can be solved by various means including the Picard-Fuchs 
equations \cite{Lerche}, rewriting the equations in terms
of a recursion relation \cite{Matone}, or by solving the
integrals directly \cite{DPKg}.

	Using similar types of arguments, these results were
generalized to other gauge groups and additions of matter
hypermultiplets \cite{SW2}-\cite{PKsoliton}.  Various other
methods of solving the resulting integrals in these generalizations
were found \cite{DPKi}-\cite{Recur}\cite{Minahan}\cite{dhokerstrong}
\cite{CalegroRecur}.

%% file: fund.tex
\setcounter{section}{0}
\setcounter{subsection}{0}
\setcounter{equation}{0}

%
%
\vspace{7mm}
\addtocounter{section}{1}
{\large {\bf \thesection.  The Seiberg-Witten Solution for Arbitrary
Classical Gauge Group $G$ }}
\medskip

We consider, $\N$=2 SUSY gauge theories with classical gauge 
groups $SU(r+1)$, $SO(2r+1)$, $Sp(2r)$ and $SO(2r)$, of rank r 
and number of colours $N_c$ = $r+1$, $2r+1$, $2r$, and $2r$
respectively.  We include $N_f$ hypermultiplets in 
the fundamental representation of the gauge group, with bare masses 
$m_j, \; j = 1, \cdots, N_f$.  We restrict to the asymptotically free
theories; this limits the hypermultiplet contents $N_f$.  
(ie. $N_f < 2N_c$ for $SU(N_c)$).
The classical vacuum expectation value of the gauge scalar $\phi$
is parameterized by complex moduli $\akb, k=1, \cdots, r$ as follows

\begin{equation}
\begin{array}{rlr}
SU(r+1) & \phi  =  diag[\ab_1, \cdots, \ab_r, \ab_{r+1}] &
\ab_1 + \cdots + \ab_r + \ab_{r+1} = 0 \\
SO(2r+1) & \phi  =  diag[\A_1, \cdots, \A_r, 0] & \\
Sp(2r) & \phi  =  diag[\ab_1, -\ab_1, \cdots, \ab_r, -\ab_r] & \\
SO(2r) & \phi  =  diag[\A_1, \cdots, \A_r] &
\A_k = 
\left({\begin{array}{cc} 0 & \ab_k \\ -\ab_k & 0 \end{array}}\right)
\end{array} 
\end{equation}
For generic $\akb$'s, the gauge symmetry is broken to $U(1)^r$ and
the dynamics is that of an Abelian Coulomb phase.  The
Wilson effective Lagrangian of the quantum theory to leading order
in the low momentum expansion in the Abelian Coulomb phase is
completely characterized by a complex analytic prepotential $\F(a)$.

The SW ansatz for determining the full 
prepotential $\F$ is based on a choice of a fibration of spectral
curves over the space of vacua, and of a meromorphic 1-form
$d\lambda$ on each of these curves.  The renormalized order
parameters $a_k$ of the theory, their duals $a_{D,k}$,
and the prepotential $\F$ are given by
\begin{eqnarray}
2\pi i a_k = \oint_{A_k} d\lambda, &
\; 2\pi i a_{D,k} = {\displaystyle \oint_{B_k}} d\lambda, & 
\; a_{D,k} = \frac{\D \F}{\D a_k} \label{eq:swanst}
\end{eqnarray}
with $A_k, B_k$ a suitable set of homology cycles on the 
spectral curves \cite{DPKg}.

For all $\N$=2 supersymmetric gauge theories based on classical gauge
groups with $N_f$ hypermultiplets in the fundamental representation of
the gauge group, the spectral curves and meromorphic 1-forms are
\begin{eqnarray}
y^2 = A^2(x) - B(x) \nonumber \\
d\lambda = \frac{x}{y}
\left(A' - \frac{A B'}{2B}\right)dx \label{eq:rsurf} 
\end{eqnarray}
where 
\begin{eqnarray}
\begin{array}{rll}
SU(r+1) \; \; & {\displaystyle A(x)=\prod_{k=1}^{r+1}(x-\akb) } & 
\; {\displaystyle B(x)=\L^2 \prod_{j=1}^{N_f}(x+m_j) } 
\end{array} \nonumber \\
\left. \begin{array}{r} 
SO(2r+1) 
\\  Sp(2r)\footnotemark
\\ SO(2r)
\end{array} \right\} \; \;
\begin{array}{ll}
{\displaystyle A(x)=x^a \prod_{k=1}^r (x^2-\akb^2) } &
\; {\displaystyle B(x)=\L^2 x^b \prod_{j=1}^{N_f}(x^2-m_j^2) }
\end{array} \label{eq:gcurves}
\end{eqnarray}
\footnotetext{For simplicity, we restrict attention here to the
Sp(2r) case with at least two massless hypermultiplets.  The cases
with one or no massless hypermultiplets may be treated accordingly
\cite{DPKg}.}
with $\L \equiv \Lambda^q$
\begin{equation}
\begin{array}{rlll}
SU(r+1) & \; \; \; q = r+1-N_f/2 & & \\
SO(2r+1) & \; \; \; q = 2r-1-N_f & \; \; a=0  & \; \; b=2 \\
Sp(2r) & \; \; \; q = 2r+2-N_f & \; \; a=2 & \; \; b=0 \\
SO(2r) & \; \; \; q = 2r-2-N_f & \; \; a=0 & \; \; b=4 \\
\end{array} \label{eq:qnum}
\end{equation}
respectively.  The spectral curves (\ref{eq:gcurves})
for $SO(2r+1)$, $Sp(2r)$ 
and $SO(2r)$ can be obtained from the $SU(2r)$ spectral curve
by a suitable restriction on the classical moduli $\akb$'s
and masses \cite{DPKg}.

For gauge theories with classical gauge groups and asymptotically free
coupling obeying the constraint $q > 0$, general
arguments based on the holomorphicity of $\F$, perturbative 
non-renormalization theorems beyond 1-loop order, the nature of instanton
corrections, and restrictions of $U(1)_R$ invariance, constrain $\F$
to have the form
\begin{eqnarray}
\F(a) & = & \frac{2q}{\pi i} \sum_{i=1}^r a_i^2 + \frac{i}{4\pi}
\left[\sum_{\alpha} (\alpha\cdot a)^2 log \frac{(\alpha\cdot a)^2}{\Lambda^2} \right. \nonumber \\ 
&& \left. - \sum_i \sum_{j=1}^{N_f} (\lambda_i \cdot a + m_j)^2 log \frac{(\lambda_i \cdot a + m_j)^2}{\Lambda^2} \right] \nonumber \\
& + & \sum_{m=1}^{\infty} 
\frac{\Lambda^{2mq}}{2m \pi i} \F^{(m)}(a)
\label{eq:prepot}
\end{eqnarray}
where $\lambda_i =\pm e_i$ for SO and Sp and 
$\lambda_i = e_i$ for SU in an orthonormal basis $\vec{e}_i$, 
and $\alpha$ are the roots of the gauge group $G$.  
(The $SU(r)$ solution requires an additional overall 
factor of $\frac{1}{2}$).

The terms on the right side are respectively the classical
prepotential, the contribution of perturbative one-loop effects, and
$m$-instanton processes contributions \footnote{The normalization of 
the instanton contributions in the present paper differs from that 
of \cite{DPKi}\cite{DPKg} by a factor $\frac{1}{4m\pi i}$ 
for $SU(N_c)$ and $\frac{1}{2m\pi i}$ for $SO(2r+1)$,
$Sp(2r)$, and $SO(2r)$.  For our purposes, it will be convenient to
use the normalization of (\ref{eq:prepot}).}.  
$\Lambda$ is the dynamically generated scale of the theory.  

\vspace{7mm}
\addtocounter{section}{1}
{\large {\bf \thesection. Renormalization Group Type Equations}}
\medskip

In \cite{DPKr}, a renormalization group type equation for the
prepotential $\F$ was derived using the SW ansatz
equations (\ref{eq:swanst}) 
\begin{equation}
\Lambda{\D\F\over\D\Lambda} = 
\frac{q}{\pi i}
\sum_{k=1}^r \akb^2 \label{eq:renormg}
\end{equation}
up to an additive term independent of $a_k$ and $\akb$ which is
physically immaterial.  (The $SU(r)$ case
requires an additional factor of $\frac{1}{2}$).  

In \cite{DPKi} an efficient algorithm was presented for 
calculating the renormalized
order parameters $a_k$ and their duals $a_{D,k}$ in terms of
the classical order parameters $\akb$ to any order
of perturbation theory in a regime where $\L$ is small
and the $\akb$'s are well-separated.
The calculation of $a_k$ starts
off from equations (\ref{eq:swanst}) and (\ref{eq:rsurf})
producing a final result 
\begin{equation}
a_k = \sum_{m=0}^{\infty}\L^{2m}{\Delta_k^{(m)}}(\bar a)
\label{eq:ak}
\end{equation}
where we set $\Delta_k^{(0)}(\ab) \equiv \ab_k$, and we have
\begin{eqnarray}
& {\displaystyle 
\Delta_k^{(m)}(\ab) = \frac{1}{2^{2m}(m!)^2} 
\left({\D\over{\D \akb}}\right)^{2m-1}  
S_k(\akb, \ab)^m,} & m \neq 0  \label{eq:del} 
\end{eqnarray}
with
\begin{equation}
\begin{array}{rl}
SU(r+1) & \; \; S_k (x, a) = {\displaystyle
\frac{\prod_{j=1}^{N_f}(x+m_j)}{\prod_{l \neq k}(x-a_l)^2} } \\
SO(2r+1) & \; \; S_k (x, a) = {\displaystyle
\frac{x^2 \prod_{j=1}^{N_f}(x^2-m_j^2)}{(x+a_k)^2 \prod_{l \neq k}(x^2-a_l^2)^2} } \\
Sp(2r)\footnotemark[1] & \; \; S_k (x, a) = {\displaystyle
\frac{\prod_{j=1}^{N_f-2}(x^2-m_j^2)}{(x+a_k)^2 \prod_{l \neq k}(x^2-a_l^2)^2} } \\
SO(2r) & \; \; S_k (x, a) = {\displaystyle
\frac{x^4 \prod_{j=1}^{N_f}(x^2-m_j^2)}{(x+a_k)^2 \prod_{l \neq k}(x^2-a_l^2)^2} } \\
\end{array} \label{eq:sfunc}
\end{equation}
and $\L$ defined as previously (\ref{eq:qnum}).

Equations (\ref{eq:renormg})(\ref{eq:ak})(\ref{eq:del})
(\ref{eq:sfunc}) suffices
to determine the prepotential $\F$ in 
terms of the renormalized order parameters $a_k$ order by order 
in powers of $\L^2$.

%
%
\vspace{7mm}
\addtocounter{section}{1}
{\large {\bf \thesection. Recursion Relation for the Prepotential $\F$}}
\medskip
	
A very direct way of deriving the form of the instanton corrections to
the prepotential $\F$ starts off from the beta function on the 
right hand side of (\ref{eq:renormg}).  Substituting the ansatz for
the prepotential (\ref{eq:prepot}) 
into the beta function (\ref{eq:renormg}), one obtains
\begin{equation}
\onesum \akb^2 = \onesum a_k^2 + 
\sum_{m=1}^{\infty} \L^{2m} \F^{(m)}(a) \label{eq:renorminst}
\end{equation}
Substituting (\ref{eq:ak}) into 
(\ref{eq:renorminst}), one obtains
\begin{eqnarray}
0 & = &  
\onesum
\left[\sum_{m=0}^{\infty}\L^{2m}\Delta_k^{(m)}(\ab) \right]^2 
- \onesum(\Delta_k^{(0)}(\ab))^2
\\ \nonumber
&&  + \sum_{m=1}^{\infty} \L^{2m} \F^{(m)}
\left(\sum_{n=0}^{\infty}\L^{2n}\Delta_k^{(n)}(\ab) \right)
\label{eq:instgen}
\end{eqnarray}
Expanding in powers of $\L^2$ in the 
last term and replacing the $\ab_k$'s with $a_k$'s, 
the $m$-th order instanton correction to the prepotential $\F$ takes on 
the form
\begin{eqnarray}
-\F^{(m)}(a) & = & \onesum
\left[\sum_{\stackrel{\scriptstyle i,j=0}{i+j=m} }^{m}\Delta_k^{(i)}(a)
\Delta_k^{(j)}(a) \right] \nonumber \\
& + & \sum_{n=1}^{m-1} \frac{1}{n!}
\sum_{\stackrel{\scriptstyle \beta_1, \cdots, \beta_{n+1} = 1}
{\beta_1 + \cdots + \beta_{n+1}=m}}^{n-1}
\sum_{\alpha_1, \cdots, \alpha_n = 1}^r
\left[ \prod_{i=1}^{n} \Delta_{\alpha_i}^{(\beta_i)}(a) \right]
\left( \prod_{j=1}^{n} \frac{\D}{\D a_{\alpha_j} } \right)
\F^{(\beta_{n+1})}(a) \nonumber \label{eq:minst}\\
\end{eqnarray}
which is a linear recursion relation for $\F^{(m)}(a)$ in terms of the lower
order instanton corrections $\F^{(m-1)}(a), \ldots, \F^{(1)}(a)$.

The intriguing part about the recursion relation (\ref{eq:minst}) 
for $\F^{(n)}(a)$ is that it is linear in 
$\F^{(n-1)}(a), \cdots, \F^{(1)}(a)$ and is valid for all classical
gauge groups with the number of hypermultiplets in the fundamental 
representation constrained by $q>0$.  Previous recursion relations
\cite{Matone} were only valid for $SU(2)$ with no hypermultiplets
and were non-linear.

\vspace{7mm}
\addtocounter{section}{1}
{\large {\bf \thesection. Instanton Expansion of the Prepotential $\F$}}
\medskip

Order by order in powers of $\L^2$, the first six instanton corrections
(\ref{eq:minst}) to the prepotential $\F$ are

\begin{eqnarray}
-\F^{(1)}(a) & = & \onesum 2\rDzero_k(a) \rDtwo_k(a) \label{eq:i2} \\
-\F^{(2)}(a) & = & \onesum
\left[ 2\rDzero_k(a) \rDfour_k(a) + (\rDtwo_k(a))^2 \right] 
+ \onesum \rDtwo_k(a) {{\D \F^{(1)}(a)}\over{\D a_k}} \\
-\F^{(3)}(a) & = & \sum_{k=1}^r
\left[ 2\rDzero_k(a) \rDsix_k(a) + 2\rDtwo_k(a)\rDfour_k(a) \right] \nonumber \\
& + & \onesum 
\left[ \rDtwo_k(a) {{\D \F^{(2)}(a)}\over{\D a_k}} \right. 
\left. + \rDfour_k(a) {{\D \F^{(1)}(a)}\over{\D a_k}} \right] \nonumber \\
& + & {1\over{2!}}\twosum
\rDtwo_k(a) \rDtwo_m(a) {{\D^2 \F^{(1)}(a)}\over{\twoden}} \\
-\F^{(4)}(a) & = & \onesum
\left[ 2\rDzero_k(a) \rDeight_k(a) + 2\rDtwo_k(a)\rDsix_k(a) + (\rDfour_k(a))^2 \right] 
\nonumber \\
& + & \onesum 
\left[ \rDtwo_k(a) {{\D \F^{(3)}(a)}\over{\D a_k}} \right. 
\left. + \rDfour_k(a) {{\D \F^{(2)}(a)}\over{\D a_k}} \right.
\left. + \rDsix_k(a) {{\D \F^{(1)}(a)}\over{\D a_k}} \right] \nonumber \\ 
& + & {1\over{2!}}\twosum
\left[ \rDtwo_k(a) \rDtwo_m(a) {{\D^2 \F^{(2)}(a)}\over{\twoden}} \right. 
+
\left. 2\rDtwo_k(a) \rDfour_m(a) {{\D^2 \F^{(1)}(a)}\over{\twoden}} \right] 
\nonumber \\
& + & {1\over{3!}}\threesum
\rDtwo_k(a) \rDtwo_m(a) \rDtwo_n(a) 
{{\D^3 \F^{(1)}(a)}\over{\threeden}} \label{eq:i8}\\
-\F^{(5)}(a) & = & \onesum
\left[ 2\rDzero_k(a) \rDten_k(a) + 2\rDtwo_k(a)\rDeight_k(a) + \right.
\left. 2\rDfour_k(a)\rDsix_k(a) \right] \nonumber \\
& + & \onesum
\left[ \rDtwo_k(a) {{\D \F^{(4)}(a)}\over{\D a_k}} \right. 
+ \rDfour_k(a) {{\D \F^{(3)}(a)}\over{\D a_k}} 
+ \rDsix_k(a) {{\D \F^{(2)}(a)}\over{\D a_k}} \nonumber \\
& + & \left. \rDeight_k(a) {{\D \F^{(1)}(a)}\over{\D a_k}} \right] 
+ {1\over{2!}}\twosum
\left[ \rDtwo_k(a) \rDtwo_m(a) {{\D^2 \F^{(3)}(a)}\over{\twoden}} \right. \nonumber \\
& + & 2\rDtwo_k(a) \rDsix_m(a) {{\D^2 \F^{(1)}(a)}\over{\twoden}} \nonumber \\
& + & \rDfour_k(a) \rDfour_m(a) {{\D^2 \F^{(1)}(a)}\over{\twoden}} 
+ \left. 2\rDtwo_k(a) \rDfour_m(a) {{\D^2 \F^{(2)}(a)}\over{\twoden}} \right]
\nonumber \\ 
& + & {1\over{3!}}\threesum
\left[ \rDtwo_k(a) \rDtwo_m(a) \rDtwo_n(a) {{\D^3 \F^{(2)}(a)}\over{\threeden}} \right.
\nonumber \\
& + & \left. 3\rDfour_k(a)\rDtwo_m(a)\rDtwo_n(a) {{\D^3 \F^{(1)}(a)}\over{\threeden}} \right]
\nonumber \\
& + & {1\over{4!}}\foursum
\rDtwo_k(a) \rDtwo_m(a) \rDtwo_n(a) \rDtwo_l(a) {{\D^4 \F^{(2)}(a)}\over{\fourden}} \\ 
-\F^{(6)}(a) & = & \onesum
\left[ 2\rDzero_k(a) \rDtwelve_k(a) + 2\rDtwo_k(a)\rDten_k(a) \right.
+ 2\rDfour_k(a)\rDeight_k(a) \nonumber \\
& + & \left. (\rDsix_k(a))^2 \right]
+ \onesum
\left[ \rDtwo_k(a) {{\D \F^{(5)}(a)}\over{\D a_k}} \right. 
+ \rDfour_k(a) {{\D \F^{(4)}(a)}\over{\D a_k}} \nonumber \\
& + & \rDsix_k(a) {{\D \F^{(3)}(a)}\over{\D a_k}} 
+ \rDeight_k(a) {{\D \F^{(2)}(a)}\over{\D a_k}}
+ \left. \rDten_k(a) {{\D \F^{(1)}(a)}\over{\D a_k}} \right] \nonumber \\
& + & {1\over{2!}}\twosum
\left[ \rDtwo_k(a) \rDtwo_m(a) {{\D^2 \F^{(4)}(a)}\over{\twoden}} \right.
+ 2\rDtwo_k(a) \rDeight_m(a) {{\D^2 \F^{(1)}(a)}\over{\twoden}} \nonumber \\
& + & \rDfour_k(a) \rDfour_m(a) {{\D^2 \F^{(2)}(a)}\over{\twoden}} 
+ 2\rDtwo_k(a) \rDfour_m(a) {{\D^2 \F^{(3)}(a)}\over{\twoden}} \nonumber \\
& + & 2\rDtwo_k(a) \rDsix_m(a) {{\D^2 \F^{(2)}(a)}\over{\twoden}} 
+ \left. 2\rDfour_k(a) \rDsix_m(a) {{\D^2 \F^{(1)}(a)}\over{\twoden}} \right]
\nonumber \\ 
& + & {1\over{3!}}\threesum
\left[ \rDtwo_k(a) \rDtwo_m(a) \rDtwo_n(a) {{\D^3 \F^{(3)}(a)}\over{\threeden}} \right. \nonumber \\
& + & 3\rDtwo_k(a) \rDtwo_m(a) \rDsix_n(a) {{\D^3 \F^{(1)}(a)}\over{\threeden}} \nonumber \\
& + & 3\rDtwo_k(a) \rDtwo_m(a) \rDfour_n(a) {{\D^3 \F^{(2)}(a)}\over{\threeden}} \nonumber \\
& + & \left. 3\rDtwo_k(a)\rDfour_m(a)\rDfour_n(a) {{\D^3 \F^{(1)}(a)}\over{\threeden}} \right]
\nonumber \\
& + & {1\over{4!}}\foursum
\left[ \rDtwo_k(a) \rDtwo_m(a) \rDtwo_n(a) \rDtwo_l(a) {{\D^4 \F^{(2)}(a)}\over{\fourden}} \right. \nonumber \\
& + & \left. 4\rDfour_k(a) \rDtwo_m(a) \rDtwo_n(a) \rDtwo_l(a) {{\D^4 \F^{(1)}(a)}\over{\fourden}} \right] \nonumber \\
& + & {1\over{5!}}\fivesum
\rDtwo_k(a) \rDtwo_m(a) \rDtwo_n(a) \rDtwo_l(a) \rDtwo_o(a) {{\D^5 \F^{(1)}(a)}\over{\fiveden}} \nonumber \\
\end{eqnarray}

A closer examination of the recursion relation 
(\ref{eq:minst}) for the prepotential $\F$ reveals
that there is always a term of the form 
\begin{equation}
2\onesum \rDzero_k(a) \Delta_k^{(n)}(a) = 
2\onesum a_k \Delta_k^{(n)}(a) \label{eq:funnyfirst}
\end{equation}
When performing explicit calculations for special cases of $N_c$ and $N_f$,
it is useful to rewrite terms of the form (\ref{eq:funnyfirst})
so that there are no $a_k$'s sitting out in front.  Using the
definition (\ref{eq:sfunc}) of $S_k(x, a)$ and performing contour
integrals in the complex plane by residue methods as in \cite{DPKi},
it can be shown that
\begin{equation}
2\onesum a_k \Delta_k^{(n)}(a) = -\frac{(2n-1)}{2^{2n-1} (n!)^2}
\onesum
\left(\frac{\D}{\D a_k}\right)^{2n-2} S_k(a_k, a)^n 
\end{equation}
up to an $a_k$ independent term that is physically immaterial for $q>0$.

\vspace{7mm}
\setcounter{subsection}{0}
\setcounter{subsubsection}{0}
\addtocounter{section}{1}
{\large {\bf \thesection. Comparison with Previous Results}}
\medskip

In order to make explicit comparisons with results in the literature,
the instanton corrections have to be rewritten in terms of
symmetric polynomials in the $a_k$'s as follows.

For $SU(2)$, the existing results in the literature have the instanton
expressions expressed in terms of
\begin{eqnarray}
a_1 & = & 2a \nonumber \\ 
a_2 & = & -2a 
\end{eqnarray}

Solving the recursion relation (\ref{eq:minst}) for the pure $SU(2)$ case, 
the explicit form for the n-th order instanton correction to the 
prepotential $\F$ was determined to be

\begin{eqnarray}
\F^{(n)}(a) =  \frac{1}{(2a)^{4n-2} } 
\sum_{j=1}^{n} \left({\begin{array}{c} 4n-3 \\ j-1 \end{array}} \right) \frac{(-1)^{j-1}}{j} \sum_{\stackrel{\scriptstyle n_1, \cdots, n_j =1}{n_1+ \cdots + n_j = n}}^{n}b_{n_1}\cdots b_{n_j} \nonumber \\
\end{eqnarray}
where
\begin{equation}
b_n = \frac{(2n-3)!!}{(n!)^2} 
\end{equation}
which agrees with previous results \cite{Lerche}\cite{DPKi}
\footnote{Our results agree exactly with those of \cite{Lerche} 
to eight instantons with the replacement 
$\Lambda^2 \rightarrow \frac{\Lambda^2}{2}$.}.

Explicit evaluations for $N_f = 0,1,2,3$ were performed, with $N_f=3$
summarized here.
\begin{eqnarray*}
\instone & = & \frac{1}{2^2 a^2}
\left[ a^2(m_1+m_2+m_3) + m_1 m_2 m_3 \right] \\
\insttwo & = & \frac{1}{2^8 a^6}
\left[ a^6 + a^4(m_1^2+m_2^2+m_3^2)- a^2(m_1^2 m_2^2 +m_1^2 m_3^2 +m_2^2 m_3^2) +  5m_1^2 m_2^2 m_3^2 \right] \\
\instthree & = & \frac{m_1 m_2 m_3}{2^{11} a^{10}}
\left[ -3a^6 + 5a^4 (m_1^2 + m_2^2 + m_3^2) - 7a^2 (m_1^2 m_2^2 + m_1^2 m_3^2 + m_2^2 m_3^2) \right. \\
&+ & \left. 9m_1^2 m_2^2 m_3^2 \right] \\
\instfour & = & \frac{1}{2^{20} a^{14}}
\left\{ a^{12} - 6a^{10}(m_1^2 + m_2^2 + m_3^2) + a^8 [5(m_1^4 +m_2^4 +m_3^4) \right. \\
& + & 100(m_1^2 m_2^2 + m_1^2 m_3^2 + m_2^2 m_3^2)] + a^6 [1176 m_1^2 m_2^2 m_3^2 \\
& - & 126(m_1^4 m_2^2 + m_1^2 m_2^4 + m_1^4 m_3^2 + m_1^2 m_3^4 + m_2^4 m_3^2 + m_2^2 m_3^4)] \\
& + & a^4 [153(m_1^4 m_2^4 + m_1^4 m_3^4 + m_2^4 m_3^4)+ 1332m_1^2 m_2^2 m_3^2(m_1^2 + m_2^2 + m_3^2)] \\
& - & 1430a^2 m_1^2 m_2^2 m_3^2 (m_1^2 m_2^2+ m_1^2 m_3^2 + m_2^2 m_3^2) 
\left. + 1469 m_1^4 m_2^4 m_3^4 \right\} \\
\instfive & = & \frac{m_1 m_2 m_3}{2^{23} a^{18}}
\left\{35 a^{12} -210a^{10} (m_1^2 + m_2^2 + m_3^2) \right. \\
& + & a^8 [207(m_1^4 + m_2^4 + m_3^2) + 
1260(m_1^2 m_2^2 + m_1^2 m_3^3 + m_2^2 m_3^2)] \\
& - & 1210a^6 (m_1^4 m_2^2 + m_1^2 m_2^4 + m_1^4 m_3^2 + m_1^2 m_3^4 
+ m_2^4 m_3^2 + m_2^2 m_3^4) \\
& + & a^4[1131(m_1^4 m_2^4 + m_1^4 m_3^4 + m_2^4 m_3^4) 
+ 5960m_1^2 m_2^2 m_3^2(m_1^2 + m_2^2 + m_3^2)] \\
& - & 5250 a^2 m_1^2 m_2^2 m_3^2 (m_1^2 m_2^2 + m_1^2 m_3^2 + m_2^2 m_3^2)
\left. + 4471 m_1^4 m_2^4 m_3^4 \right\} \\
\instsix & = & \frac{1}{2^{29}a^{22} }
\left\{5a^{16}(m_1^2 + m_2^2 + m_3^2) - a^{14}[210(m_1^2 m_2^2 + m_1^2 m_3^2 + m_2^2 m_3^2) \right. \\
& + & 14(m_1^4 + m_2^4 + m_3^4)] + a^{12}[9(m_1^6 + m_2^6 + m_3^6) + 
6507 m_1^2 m_2^2 m_3^2 \\
& + & 801(m_1^4 m_2^2 + m_1^2 m_2^4 + m_1^4 m_3^2 + m_1^2 m_3^4 + m_2^4 m_3^2
+ m_2^2 m_3^4)] \\
& - & a^{10}[660(m_1^6 m_2^2 + m_1^2 m_2^6 + m_1^6 m_3^2 + m_1^2 m_3^6 
+ m_2^6 m_3^2 + m_2^2 m_3^6) \\
& + & 330(m_1^4 m_2^4 + m_1^4 m_3^4 + m_2^4 m_3^4)
+ 24420m_1^2 m_2^2 m_3^2 (m_1^2 + m_2^2 + m_3^2)] \\
& + & a^8[2769(m_1^6 m_2^4 + m_1^4 m_2^6 + m_1^6 m_3^4 + m_1^4 m_3^6 
+ m_2^6 m_3^4 + m_2^4 m_3^6) \\
& + & m_1^2 m_2^2 m_3^2 (19851(m_1^4 + m_2^4 + m_3^4) 
+ 87945(m_1^2 m_2^2 + m_1^2 m_3^2 + m_2^2 m_3^2))] \\ 
& - & a^6[295050 m_1^4 m_2^4 m_3^4 + 2310(m_1^2 m_2^2 + m_1^2 m_3^2
+ m_2^2 m_3^2) \\
& + & 69510m_1^2 m_2^2 m_3^2 (m_1^4 m_2^2 + m_1^2 m_2^4 + m_1^4 m_3^2 
+ m_1^2 m_3^4 + m_2^4 m_3^2 + m_2^2 m_3^4)] \\
& + & a^4m_1^2 m_2^2 m_3^2 [53839(m_1^4 m_2^4 + m_1^4 m_3^4 + m_2^4 m_3^4) \\
&+& 224485m_1^2 m_2^2 m_3^2 (m_1^2+m_2^2+m_3^2)] \\
& - & 166896a^2 m_1^4 m_2^4 m_3^4 (m_1^2 m_2^2 +m_1^2 m_3^2 +m_2^2 m_3^2)
+ \left. 121191 m_1^6 m_2^6 m_3^6 \right\}
\end{eqnarray*}
\medskip

A check of the hypermultiplet decoupling limits of 
the $N_f=3$ instanton corrections,
by letting $\Lambda_3 m_3 = \Lambda_2^2$ and
sending $m_3 \rightarrow \infty$, reproduces the $N_f=2$ results.
A further decoupling of a second hypermultiplet, by letting 
$\Lambda_2 m_2 = \Lambda_1^2$ and sending $m_c \rightarrow \infty$,
reproduces the $N_f=1$ results.
Comparison with results in the literature \cite{Ohta1}\cite{Ohta2}\cite{DPKi}
\cite{SpanW1}\cite{SpanW2}
show an agreement to four instantons up to a redefinition of the
$\akb$'s as discussed in \cite{DPKi}.

For $SU(3)$, the existing results in the literature have the instanton
corrections expressed in terms of the invariant $SU(3)$ symmetric 
polynomials $u,v$ and the discriminant $\Delta$
\begin{eqnarray}
u & = & - a_1 a_2 - a_1 a_3 - a_2 a_3 \nonumber \\
v & = & a_1 a_2 a_3 \nonumber \\
\Delta & = & 4u^3 - 27 v^3
\end{eqnarray}
and the $p$-th symmetric mass polynomials
\begin{equation}
t_p(m) = \sum_{j_1 < \cdots < j_p} m_{j_1}\cdots m_{j_p}
\end{equation}

Explicit evaluations for $N_f = 0,1,2,3,4,5$ were performed, and are 
summarized here for $N_f=0$.
\begin{eqnarray*}
\instone & = & \sL^6 \frac{3u}{\Delta} \\
\insttwo & = & \frac{\sL^{12} u}{16} 
\left[ \frac{10935 v^2}{\Delta^3} + \frac{153}{\Delta^2} \right] \\
\instthree & = & \frac{3\sL^{18} u}{16}
\left[ \frac{4782969 v^4}{2\Delta^5} + \frac{161109 v^2}{2\Delta^4} \right.
\left. + \frac{385}{\Delta^3} \right] \\
\instfour & = & \frac{\sL^{24} u}{4096}
\left[ \frac{1707362095023v^6}{\Delta^7} + \frac{91216001799 v^4}{\Delta^6} \right.
+ \frac{1254600981 v^2}{\Delta^5} \\
&& \left. + \frac{3048885}{\Delta^4} \right] \\
\instfive & = & \frac{5\sL^{30} u}{4096}
\left[ \frac{3788227372819653 v^8}{10\Delta^9} + \frac{277223767370307 v^6}{10 \Delta^8} \right. \\
&& +\frac{6447389599341 v^4}{10\Delta^7} + \frac{50110037721 v^2}{10\Delta^6}
\left. + \frac{7400133}{\Delta^5} \right] \\
\instsix & = & \frac{3\sL^{36}u}{65535}
\left[ \frac{24952152189682606959 v^{10} }{2\Delta^{11} } \right.
+ \frac{2319087386959542567 v^8}{2\Delta^{10} } \\
&& + \frac{38185135433846901 v^6}{\Delta^9}
+ \frac{525166021552761 v^4}{\Delta^8} \\
&& + \frac{5323867298775 v^2}{2\Delta^7}
+\left. \frac{5295230391}{2\Delta^6} \right] 
\end{eqnarray*}
A check of successive hypermultiplet decoupling limits of the 
$N_f=5$ instanton corrections reproduces 
all of the $N_f < 5$ cases accordingly.
Comparison with results in the literature 
\cite{Lerche}\cite{DPKi}\cite{Ewen1}\cite{Ewen2}
\cite{SpanW1}\cite{SpanW2}
show an agreement to three instantons up to a redefinition of the
$\akb$'s as discussed in \cite{DPKi}.

\vspace{7mm}

\setcounter{subsection}{0}

\vspace{7mm}
\addtocounter{section}{1}
{\large {\bf \thesection.  Classical Moduli in Terms of Quantum Moduli} }
\medskip

Another way of evaluating the beta function (\ref{eq:renormg}) 
of the prepotential $\F$ involves inverting (\ref{eq:ak})
to get
\begin{equation}
\akb \equiv a_k + 
\sum_{m=1}^{\infty} \L^{2m} \beta_k^{(m)}(a) \label{eq:betas}
\end{equation}	
where the $\beta_k(a)$'s are functions of the renormalized order
parameters $a_k$. 

A very direct way of deriving the form of the $\beta_k(a)$'s
involves starting off with (\ref{eq:betas}) and
substituting in equation (\ref{eq:ak}) to get
\begin{eqnarray}
0 = \sum_{m=1}^{\infty}\L^{2m}{\Delta_i^{(m)}}(\bar a) +
\sum_{m=1}^{\infty}\L^{2m}{\beta_i^{(m)}}
\left( \sum_{m=0}^{\infty}\L^{2m}{\Delta_k^{(m)}}(\bar a) \right)
\label{eq:generate}
\end{eqnarray}
Expanding in powers of $\L^2$ in the second term 
and replacing the $\akb$'s with $a_k$'s, one obtains
\begin{eqnarray}
-\beta_k^{(m)}(a) & = & \Delta_k^{(m)}(a) \nonumber \\
& + & \sum_{n=1}^{m-1} \frac{1}{n!}
\sum_{\stackrel{\scriptstyle \beta_1, \cdots, \beta_{n+1} = 1}
{\beta_1 + \cdots + \beta_{n+1}=m}}^{n-1}
\sum_{\alpha_1, \cdots, \alpha_n = 1}^r
\left[ \prod_{i=1}^{n} \Delta_{\alpha_i}^{(\beta_i)}(a) \right]
\left( \prod_{j=1}^{n} \frac{\D}{\D a_{\alpha_j} } \right)
\beta_k^{(\beta_{n+1})}(a) \nonumber \label{eq:bk}\\
\end{eqnarray}
Order by order in powers of $\L^2$, the first few $\beta_k(a)$'s are 
\begin{eqnarray*}
-\bktwo & = & \rDtwo_k(a)  \\
-\bkfour & = & \rDfour_k(a) + \sum_{l=1}^r \rDtwo_l(a)
{{\D \bktwo}\over{\D a_l}} \\
-\bksix & = & \rDsix_k(a) + \sum_{l=1}^r
\left[ \rDtwo_l(a) {{\D \bkfour}\over{\D a_l}} \right.
\left. + \rDfour_l(a) {{\D \bktwo}\over{\D a_l}} \right] \\
& + & {1\over{2!}}\sum_{l,m=1}^r \rDtwo_l(a) \rDtwo_m(a)
{ {\D^2 \bktwo}\over {\D a_l \D a_m} } \\
-\bkeight & = & \rDeight_k(a) + \sum_{l=1}^r
\left[ \rDtwo_l(a) {{\D \bksix}\over{\D a_l}} \right.
\left. + \rDfour_l(a) {{\D \bkfour}\over{\D a_l}} \right. \\
& + & \left. \rDsix_l(a) {{\D \bktwo}\over{\D a_l}} \right]
+ {1\over{2!}}\sum_{l,m=1}^r 
\left[\rDtwo_l(a) \rDtwo_m(a) {{\D^2 \bkfour}\over{\D a_l \D a_m}} \right. \\
& + & \left. 2\rDtwo_l(a) \rDfour_m(a) {{\D^2 \bktwo}\over{\D a_l \D a_m}} \right] \\
& + & {1\over{3!}}\sum_{l,m,n=1}^r
\rDtwo_l(a) \rDtwo_m(a) \rDtwo_n(a)
{ {\D^3 \bktwo}\over{\D a_l \D a_m \D a_n} }
\end{eqnarray*}
Substituting (\ref{eq:bk}) into (\ref{eq:renorminst})
reproduces the instanton corrections to the prepotential
(\ref{eq:minst}) order by order in $\L^2$.

%% file: adj.tex
%
%
\def\k{{\bf k}}
\def\a{{\bf a}}

\setcounter{section}{0}
\setcounter{subsection}{0}
\setcounter{equation}{0}

\addtocounter{section}{1}
{\large {\bf \thesection.  The Seiberg-Witten Solution for Super Yang-Mills with One Adjoint Hypermultiplet}}
\medskip

For supersymmetric Yang-Mills theories with an asymptotic free coupling 
and one adjoint hypermultiplet in the adjoint representation of a classical
gauge group, general arguments based on the holomorphicity of $\F$, 
perturbative nonrenormalization theorems beyond 1-loop order, the nature
of instanton corrections, and restrictions of $U(1)_R$ invariance constrain
$\F$ to have the form
\begin{eqnarray}
\F(a) & = & \frac{\tau}{2} \sum_{i=1}^r a_i^2 -\frac{1}{8\pi i} 
\sum_{\alpha \in {\cal R(G)} } \left\{(\alpha \cdot a)^2 log (\alpha \cdot a)^2 
\right.
\nonumber \\
& & \left. - (\alpha \cdot a+m)^2 log(\alpha \cdot a+m)^2\right\} 
+ \sum_{n=1}^\infty \frac{q^n}{2\pi n i}\F^{(n)}(a) 
\label{eq:pertpre}
\end{eqnarray}
where $\alpha$ are the roots of the gauge group $\cal{G}$.  
For $SU(N)$, the traceless condition $\sum_{i=1}^N a_i = 0$ is imposed.

The SW ansatz for determining the full prepotential $\F$ is based on 
a choice of a fibration of spectral curves over the space of vacua,
and of a meromorphic 1-form $d\lambda$ on each of these curves.
The renormalized order parameters $a_k$ of the theory, their duals
$a_{D,k}$, and the prepotential $\F$ are given by
\begin{eqnarray}
2\pi i a_k = \oint_{A_k} d \lambda, &
\; 2\pi i a_{D, k} = {\displaystyle \oint_{B_k} } d \lambda, &
\; a_{D,k} = \frac{\D \F}{\D a_k}
\label{eq:sw}
\end{eqnarray}
with $A_k, B_k$ a suitable set of homology cycles on the spectral curves.

For $\N=2$ supersymmetric gauge theories with gauge group $SU(N)$ and one
hypermultiplet in the adjoint representation, a
convenient parameterization for the spectral curves and meromorphic 1-forms 
is the Calogero-Moser case of \cite{Calegro}
\begin{eqnarray}
f(k-\frac{m}{2}, z) = 0, & d \lambda = k dz \label{eq:spectral}
\end{eqnarray}
where
\begin{eqnarray}
f(k,z)  & = & {\displaystyle
\frac{1}{\vartheta_1 (\frac{z}{2 \omega_1} | \tau)}\sum_{n=0}^N
\frac{1}{n!} \frac{\D^n}{\D z^n} \vartheta_1 (\frac{z}{2 \omega_1} | \tau)
(-m \frac{\D}{\D k})^n H(k | \k) } \\ \label{eq:tspec}
H(x | \k) & = & \prod^N_{j=1}(x-k_j) \equiv (x-k_i) H_i(x | \k) \\
\vartheta_1 (z | \tau) & = & \sum_{n \in Z} (-1)^n q^{(n+1/2)^2/2} 
e^{2\pi i(2n+1)z} \label{eq:theta}
\end{eqnarray}
along with a corresponding basis of 
$A_k, B_k$ homology cycles as described in \cite{Calegro}.
This particular choice of parameterization for the spectral curves has 
the geometry of a foliation over a base torus $\Sigma$, where the complex
modulus $\tau$ of the torus $\Sigma$ is related to the gauge coupling
$g$ and the $\theta$-angle of the gauge theory by
\begin{eqnarray}
\tau = \frac{\theta}{2\pi} + \frac{4\pi i}{g^2}
\label{eq:torus}
\end{eqnarray}

Substituting (\ref{eq:theta}) into
(\ref{eq:spectral}) produces a
simplified form for the spectral curves
\begin{eqnarray}
\sum_{n \in Z} (-1)^n q^{\frac{1}{2}n(n-1)}e^{nz} H(k-mn | \k) = 0
\label{eq:simspec}
\end{eqnarray}

Substituting (\ref{eq:simspec}) into (\ref{eq:sw}) using the 
Calogero-Moser parameterization (\ref{eq:spectral}) for the 1-form and
performing a weak coupling expansion in powers of $q$ similar to 
the methods in \cite{Calegro}, the integral for the
quantum order parameters $a_i$'s in terms of the classical
order parameters $k_i$'s were calculated order by order in $q$
producing a simplified expression of
\begin{eqnarray}
a_i &  = &  k_i + \sum_{n=1}^\infty q^n \Delta^{(n)}_i (k)
\nonumber
\end{eqnarray}
where
\begin{eqnarray}
\sum_{n=1}^\infty q^n \Delta^{(n)}_i (k) & = & \sum_{j=2}^\infty
\;
\sum^\infty_{\stackrel{\alpha_1, \cdots, \alpha_j = -\infty, \neq 0}
{\alpha_1 + \cdots + \alpha_j = 0} }
\frac{(-1)^j}{j!} (\frac{\D}{\D k_i})^{j-1} 
\prod_{l=1}^j \left[\frac{H(k_i - \alpha_l m | \k)}{H_i (k_i | \k)} q^{\alpha_l^2/2}
\right] \nonumber \\ \label{eq:qorder}
\end{eqnarray}

The first few $\Delta_i$'s are

\begin{eqnarray}
\Delta_i^{(1)}(a) & = &
\frac{\D}{\D a_i}
\left[ \frac{H(a_i-m | \a)H(a_i+m | \a)}{H_i (a_i | \a)^2} \right]
\nonumber \\
\Delta_i^{(2)}(a) & = &
\frac{1}{4}
\frac{\D^3}{\D a_i^3} 
\left[\frac{H(a_i-m | \a)H(a_i+m | \a)}{H_i (a_i | \a)^2}\right]^2
\nonumber \\
\Delta_i^{(3)}(a) & = &
\frac{1}{36} \frac{\D^5}{\D a_i^5} 
\left[\frac{H(a_i-m | \a)H(a_i+m | \a)}{H_i (a_i | \a)^2}\right]^3
\nonumber \\
& - & \frac{1}{2}
\frac{\D^2}{\D a_i^2}
\left[ \frac{H(a_i-2m| \a) H(a_i+m | \a)^2}{H_i (a_i | \a)^3} \right]
\nonumber \\
&-& \frac{1}{2}
\frac{\D^2}{\D a_i^2}
\left[ \frac{H(a_i-m | \a)^2 H(a_i+2m | \a)}{H_i (a_i | \a)^3} \right]
\nonumber \\
\Delta_i^{(4)}(a) & = & 
\frac{1}{576} \frac{\D^7}{\D a_i^7} 
\left[\frac{H(a_i-m | \a)H(a_i+m | \a)}{H_i (a_i | \a)^2}\right]^4
\nonumber \\
&+& \frac{\D}{\D a_i} 
\left[ \frac{H(a_i-2m | \a) H(a_i+2m | \a)}{H_i (a_i | \a)^2} \right]
\nonumber \\
& - & \frac{1}{6}
\frac{\D^4}{\D a_i^4}
\left[\frac{H(a_i-2m | \a) H(a_i-m | \a) H(a_i+m | \a)^3}{H_i (a_i | \a)^5}\right] 
\nonumber \\
& - & \frac{1}{6}
\frac{\D^4}{\D a_i^4}
\left[\frac{H(a_i-m | \a)^3 H(a_i+m | \a)H(a_i+2m | \a)}{H_i (a_i | \a)^5}\right]
\nonumber 
\end{eqnarray}

This result can be derived in a more transparent manner by rewriting 
the spectral curves (\ref{eq:simspec}) as
\begin{eqnarray}
k & \equiv & k_i + F_i (k)
\end{eqnarray}
where
\begin{eqnarray}
F_i (k) & = &   \sum_{n \in Z, \neq 0} (-1)^{n+1}
q^{\frac{1}{2} n(n-1)} e^{nz} \frac{H(k-nm | \k)}{H_i (k | \k)}
\end{eqnarray}
An iterative solution expanded around $k = k_i$ to all orders 
in small q is given by
\begin{eqnarray}
k = k_i + \sum_{n=1}^\infty y_n, & &  
y_n = \frac{1}{n!} \frac{\D^{n-1} }{\D k^{n-1} } F_i^n (k) |_{k= k_i}
\label{eq:iterative}
\end{eqnarray}

In a similar manner as in \cite{Calegro}, $z$ is substituted 
with $w =  e^z$ in the the SW differential (\ref{eq:spectral}),
\begin{eqnarray}
dz = \frac{dw}{w}, && 
d \lambda = k dz  = k \frac{dw}{w} 
\label{eq:subswdiff}
\end{eqnarray}
along with the iterative solution (\ref{eq:iterative}).  
Performing the integral around the
appropriate $A_i$ cycle corresponding to $k_i$ as prescribed in
\cite{Calegro}, reproduces (\ref{eq:qorder}) \cite{Pdhoker}.

\vspace{7mm}
\addtocounter{section}{1}
{\large {\bf \thesection.  Renormalization Group Type Equations}}
\medskip

In \cite{Calegro}, a renormalization group type equation for the   
prepotential $\F$ was derived
\begin{equation}
{\D\F\over\D\tau} = {1\over{4\pi i}} \sum_{j=1}^r \oint_{A_j} k^2 dz
\label{eq:rg}
\end{equation}
up to an additive term independent of $a_i$ and $k_i$ which is 
physically immaterial.

Substituting the $SU(N)$ spectral curves (\ref{eq:simspec}) 
into (\ref{eq:rg}) using the Calogero-Moser
parameterization (\ref{eq:spectral}) for the 1-form and solving
the integral in the weak coupling limit of small $q$
gives the renormalization group like equation for the prepotential
$\F$ in terms of the classical order parameters $k_i$'s

\begin{eqnarray}
{\D\F\over\D\tau} & = &  \frac{1}{2} \sum_{i=1}^r k_i^2
+ \sum_{i=1}^r \sum_{n=1}^\infty q^n k_i \Delta_i^{(n)}(k)
+ \sum_{i=1}^r \sum_{n=1}^\infty q^n \Omega_i^{(n)}(k)
\nonumber
\end{eqnarray}
where
\begin{eqnarray}
\sum_{n=1}^\infty q^n \Omega^{(n)}_i (k) & = & \sum_{j=2}^\infty
\;
\sum^\infty_{\stackrel{\alpha_1, \cdots, \alpha_j = -\infty, \neq 0}
{\alpha_1 + \cdots + \alpha_j = 0} }
\frac{(-1)^j}{j(j-2)!} (\frac{\D}{\D k_i})^{j-2} 
\prod_{l=1}^j \left[ \frac{H(k_i - \alpha_l m | \k)}{H_i (k_i | \k)} q^{\alpha_l^2/2}
\right] \nonumber \\
& &
\label{eq:renormgr}
\end{eqnarray}

The first few $\Omega_i$'s are

\begin{eqnarray}
\Omega_i^{(1)}(a) & = &
\frac{H(a_i-m | \a)H(a_i+m | \a)}{H_i (a_i | \a)^2}
\nonumber \\
\Omega_i^{(2)}(a) & = &
\frac{3}{4} \frac{\D^2}{\D a_i^2} 
\left[\frac{H(a_i-m | \a)H(a_i+m | \a)}{H_i (a_i | \a)^2}\right]^2
\nonumber \\
\Omega_i^{(3)}(a) & = &
\frac{5}{36} \frac{\D^4}{\D a_i^4} 
\left[\frac{H(a_i-m | \a)H(a_i+m | \a)}{H_i (a_i | \a)^2}\right]^3
\nonumber \\
&-& \frac{\D}{\D a_i}
\left[ \frac{H(a_i-2m| \a) H(a_i+m | \a)^2}{H_i (a_i | \a)^3} \right]
\nonumber \\
&-& \frac{\D}{\D a_i}
\left [\frac{H(a_i-m | \a)^2 H(a_i+2m | \a)}{H_i (a_i | \a)^3} \right]
\nonumber \\
\Omega_i^{(4)}(a) & = & 
\frac{7}{576} \frac{\D^6}{\D a_i^6} 
\left[\frac{H(a_i-m | \a)H(a_i+m | \a)}{H_i (a_i | \a)^2}\right]^4
\nonumber \\
&+& \frac{H(a_i-2m | \a) H(a_i+2m | \a)}{H_i (a_i | \a)^2}
\nonumber \\
& - & \frac{2}{3}
\frac{\D^3}{\D a_i^3}
\left[\frac{H(a_i-2m | \a) H(a_i-m | \a) H(a_i+m | \a)^3}{H_i (a_i | \a)^5}\right] 
\nonumber \\
& - & \frac{2}{3}
\frac{\D^3}{\D a_i^3}
\left[\frac{H(a_i-m | \a)^3 H(a_i+m | \a)H(a_i+2m | \a)}{H_i (a_i | \a)^5}\right]
\nonumber 
\end{eqnarray}

%
%
\vspace{7mm}
\addtocounter{section}{1}
{\large {\bf \thesection. Recursion Relations for the Prepotential $\F$}}
\medskip

In \cite{Recur}, an efficient algorithm for deriving a set of recursion
relations for the instanton corrections was discovered.  Using similar
methods as \cite{Recur}, a similar set of recursion relations for
the prepotential $\F$ was determined.

A very direct way of deriving the form of the instanton corrections to the
prepotential $\F$ involves
substituting (\ref{eq:pertpre}) and (\ref{eq:qorder}) into 
(\ref{eq:renormgr}) to get
\begin{eqnarray}
\sum_{n=1}^\infty q^n \F^{(n)}(a) & = & 
\sum_{i=1}^r \sum_{n=1}^\infty q^n \Omega^{(n)}_i (k) 
- \frac{1}{2}
\sum_{i=1}^r \left[\sum_{n=1}^\infty q^n \Delta^{(n)}_i (k)\right]^2
\label{eq:istart}
\end{eqnarray}
Then (\ref{eq:qorder}) is substituted into (\ref{eq:istart})
and expanded in powers of $q$, replacing the $k_i$'s with $a_i$'s.
The n-th order instanton correction to the prepotential $\F$ 
finally takes on the form
\begin{eqnarray}
\F^{(n)}(a) & = & \sum_{i=1}^r \Omega^{(n)}_i (a) 
-\frac{1}{2} \sum_{i=1}^r \sum_{\stackrel{j,l=1}{j+l=n}}^n 
\Delta^{(j)}_i(a) \Delta^{(l)}_i(a)
\nonumber \\
& - & \sum_{i=1}^{n-1} \frac{1}{i!}
\sum_{\stackrel{\beta_1, \cdots, \beta_{i+1} = 1}{\beta_1 + \cdots + \beta_{i+1} = n} }^{i-1}
\sum_{\alpha_1, \cdots, \alpha_i =1}^r 
\left[\prod_{j=1}^i \Delta_{\alpha_j}^{(\beta_j)}(a)\right]
\left(\prod_{l=1}^i \frac{\D}{\D a_{\alpha_l} }\right) \F^{(\beta_{i+1})}(a)
\nonumber \\ \label{eq:prepoten}
\end{eqnarray}

The first few $\F^{(n)}(a)$'s are

\begin{eqnarray}
\F^{(1)}(a) & = & \sum_{i=1}^r \Omega_i^{(1)}(a)
\nonumber \\
\F^{(2)}(a) & = & \sum_{i=1}^r \Omega_i^{(2)}(a)
-\frac{1}{2} \sum_{i=1}^r (\Delta_i^{(1)}(a))^2
-\sum_{i=1}^r \Delta_i^{(1)}(a) \frac{\D \F^{(1)}(a)}{\D a_i}
\nonumber \\
\F^{(3)}(a) & = & \sum_{i=1}^r \Omega_i^{(3)}(a)
-\frac{1}{2} \sum_{i=1}^r \left[2\Delta_i^{(1)}(a) \Delta_i^{(2)}(a)\right]
\nonumber \\
& - & \sum_{i=1}^r \left[
\Delta_i^{(1)}(a) \frac{\D \F^{(2)}(a)}{\D a_i}
+ \Delta_i^{(2)}(a) \frac{\D \F^{(1)}(a)}{\D a_i} \right]
\nonumber \\
&-& \frac{1}{2!} \sum_{i,j=1}^r 
\Delta_i^{(1)}(a) \Delta_j^{(1)}(a) \frac{\D^2 \F^{(1)}(a)}{\D a_i \D a_j}
\nonumber \\
\F^{(4)}(a) & = & \sum_{i=1}^r \Omega_i^{(4)}(a)
-\frac{1}{2} \sum_{i=1}^r \left[2\Delta_i^{(1)}(a) \Delta_i^{(3)}(a)
+(\Delta_i^{(2)}(a))^2 \right]
\nonumber \\
& - & \sum_{i=1}^r \left[
\Delta_i^{(1)}(a) \frac{\D \F^{(3)}(a)}{\D a_i}
+ \Delta_i^{(2)}(a) \frac{\D \F^{(2)}(a)}{\D a_i}
+ \Delta_i^{(3)}(a) \frac{\D \F^{(1)}(a)}{\D a_i} \right]
\nonumber \\
&-& \frac{1}{2!} \sum_{i,j=1}^r \left[
\Delta_i^{(1)}(a) \Delta_j^{(1)}(a) \frac{\D^2 \F^{(2)}(a)}{\D a_i \D a_j}
+ 2\Delta_i^{(1)}(a) \Delta_j^{(2)}(a) \frac{\D^2 \F^{(1)}(a)}{\D a_i \D a_j}
\right] \nonumber \\
&-& \frac{1}{3!} \sum_{i,j,k=1}^r 
\Delta_i^{(1)}(a) \Delta_j^{(1)}(a) \Delta_k^{(1)}(a)
\frac{\D^3 \F^{(1)}(a)}{\D a_i \D a_j \D a_k}
\end{eqnarray}

\vspace{7mm}
\addtocounter{section}{1}
{\large {\bf \thesection.  Comparison With Previous Results}}
\medskip

In the limit the full hypermultiplet is decoupled with
$\tau \rightarrow \infty, m \rightarrow \infty$ while keeping
constant the parameters $k_i$ and $\Lambda$:
\begin{eqnarray}
\Lambda^{2N} = (-1)^N m^{2N} q & &  q = e^{2\pi i \tau}
\end{eqnarray} 
equations (\ref{eq:qorder}) and (\ref{eq:renormgr}) break down to their
corresponding equations in the pure $SU(N)$ gauge theory cases 
\cite{DPKg}\cite{Recur}\cite{DPKr}.

In the $\N$=4 limit where $m \rightarrow 0$, all the $\Omega_i$ and
$\Delta_i$ terms in (\ref{eq:qorder}) and (\ref{eq:renormgr}) vanish
and reproduces the expected prepotential
\begin{eqnarray}
\F (a) = \frac{\tau}{2}\sum_{i=1}^N a_i^2
\end{eqnarray}
for $SU(N)$.

For $SU(2)$, the existing results in the literature have the instanton
expressed in term of
\begin{eqnarray}
a_1 = a, & a_2 = -a
\end{eqnarray}
Explicit evaluations of the first three instanton corrections 
are

\begin{eqnarray}
\F^{(1)}(a) & = & \frac{m^4}{2a^2} \nonumber \\
\frac{\F^{(2)}(a)}{2} & = & -\frac{9 m^6}{16 a^4} 
+ \frac{5 m^8}{64 a^6} \nonumber \\
\frac{\F^{(3)}(a)}{3} & = & \frac{m^6}{a^4} + \frac{25 m^8}{48 a^6}
-\frac{67 m^{10}}{192 a^8} + \frac{3 m^{12}}{64 a^{10}} 
\label{eq:su2result}
\end{eqnarray}
which disagree with results in the literature beyond one instanton
\cite{Minahan}
\footnote{
In the limit of decoupling
the full $SU(2)$ adjoint hypermultiplet in \cite{Minahan}, 
there is a discrephency of a factor
$\frac{1}{2}$ with the pure $SU(2)$ results of \cite{Lerche}\cite{Recur}.}
, but agrees in the limit where the full
hypermultiplet decouples \cite{Recur}.  It turns out that peforming the 
Seiberg-Witten elliptic function calculation in \cite{Dorey} 
to higher instanton orders reproduces the
instanton calculations of \cite{Minahan}.  

On the other hand, the $SU(2)$ spectral curve from the 
Calogero-Moser construction (\ref{eq:spectral}) can be explicitly 
shown to be equivalent to the $SU(2)$ mass deformed $\N$=4 
spectral curve construction \cite{SW2} up to reparameterizations
of the classical order parameters $k_i$'s.  This spectral curve
forms a crucial part of the elliptic function calculation in 
\cite{Dorey}.

One possible problem with the elliptic function calculation in \cite{Dorey}
is the assumption of Matone's relation 
\cite{Matone}\cite{Sonnen} holding in the 
presence of an adjoint hypermultiplet.  Generalizations of Matone's relation
for classes of $\N$=2 SUSY gauge theories with fundamental hypermultiplets 
was proven in general \cite{DPKr} and corresponds to a renormalization
group type of equation for the prepotential $\F$.  For the case of one adjoint
hypermultiplet, a renormalization group equation for the prepotential $\F$ 
was derived \cite{Calegro} and calculated to all orders (\ref{eq:renormgr})
which differs greatly from Matone's relation and 
\cite{Sonnen}\cite{DPKr}, but
agrees with the latter cases in the limit the full adjoint hypermultiplet is
decoupled \cite{Calegro}.

\vspace{7mm}
\addtocounter{section}{1}
{\large {\bf \thesection. S-Duality Properties}}
\medskip

A closer examination of the Calogero-Moser parameterization of the
Seiberg-Witten spectral curves and 1-form (\ref{eq:spectral}) and
(\ref{eq:tspec}) reveals there's an implicit S-duality present.  

Using the transformation property of the theta functions
\begin{eqnarray}
\vartheta_1 \left(\frac{z}{\tau} | -\frac{1}{\tau} \right)
= \sqrt{-i\tau} \exp \left( \frac{iz^2}{\pi\tau} \right) 
\vartheta_1 (z | \tau)
\end{eqnarray}  
and substituting it into (\ref{eq:spectral}) and (\ref{eq:tspec}) shows
explicitly that the form of the spectral curves and 1-form are indeed invariant
under S-duality transformations up to reparameterizations of the
classical order parameters $k_i$'s.  Correspondingly, the roles of the
A and B cycles in the Seiberg-Witten ansatz (\ref{eq:sw}) are interchanged
under S-duality transformations.  

With this explicit S-duality, the corresponding weakly coupled "dual" 
theory in the magnetic sector of the theory 
expanded around a small "dual" coupling constant
\begin{eqnarray}
q_D = e^{2\pi i \tau_D} & \tau_D = -\frac{1}{\tau}
& \tau \rightarrow i0^+ 
\end{eqnarray}
will have a corresponding "dual" prepotential $\F_D(a_D)$ identical in
form to the prepotential $\F(a)$ in 
(\ref{eq:pertpre}) and (\ref{eq:prepoten}) with the
corresponding substitutions of the coupling constant and 
quantum order parameters to their "dual" counterparts
\begin{eqnarray}
q \rightarrow q_D & & a_i \rightarrow a_{D,i} 
\end{eqnarray}
respectively.  This can be interpreted as a non-perturbative expansion
of the theory, where the dynamics of the strongly coupled
regime in the electric
sector of the theory is described by the dynamics of a corresponding weakly 
coupled "dual" theory in the magnetic sector of the same theory.  (Other
strong coupling expansions in the same spirit were performed in 
\cite{Lerche}\cite{dhokerstrong}\cite{EdelStrong}).

Considering there are claims that the Calogero-Moser system 
can be constructed explicitly from the Hitchin system \cite{Markman}, this
S-duality is like a realization of the Donagi-Witten construction of
Seiberg-Witten theory using the Hitchin system
\cite{Donagi} where an underlying S-duality and general $SL(2, Z)$
symmetry is built 
into the geometry of the foliation over a base torus $\Sigma$
construction (\ref{eq:torus}) from the start.
In the prepotential calculations performed around small 
coupling $q$ or $q_D$, the S-duality is explicitly broken while the
underlying spectral curve (\ref{eq:spectral}) is invariant 
under S-duality and in general an $SL(2, Z)$ symmetry 
\cite{SW2}\cite{Donagi}.

\vspace{7mm}
\addtocounter{section}{1}
{\large {\bf \thesection. Generalizations to Other Gauge Groups}}
\medskip

Generalizations of the $SU(N)$ Calogero-Moser integrable system were
investigated in \cite{DHokerPhong1}\cite{DHokerPhong2} 
for various cases of twisted and untwisted
gauge groups, but stopped short of producing parameterizations suitable
for use as Seiberg-Witten spectral curves.
Possible parameterizations to general untwisted classical gauge groups
can be conjectured starting from the $SU(N)$ spectral curves and placing
appropriate constraints such that decouplings of the full adjoint
hypermultiplet reproduce the pure gauge theory results 
\cite{Brandhuber}\cite{Daniel}\cite{Hanany}\cite{DHokerGen}.

In the spirit of \cite{Hanany}\cite{DHokerGen}, one possibility is to 
replace the $H(k)$ polynomial with
\begin{eqnarray}
H(x | \k ) &\rightarrow &
H(x | \k)  =  \prod^N_{j=1}(x^2-k^2_j) \equiv (x-k_i)(x+k_i) H_i(x | \k) 
\end{eqnarray}
in the Calogero-Moser parameterization of the SW spectral curves 
(\ref{eq:simspec}).

The appropriate limits for full hypermultiplet decoupling are 
$\tau \rightarrow \infty, m \rightarrow \infty$ while keeping
constant the parameters $k_i$ and $\Lambda$:
\begin{eqnarray}
SO(2r) & & \Lambda^{4r-4} \equiv m^{4r-4}q \nonumber \\
SO(2r+1) & & \Lambda^{4r-2} \equiv m^{4r-2}q \nonumber \\
SO(2r) & & \Lambda^{4r+4} \equiv m^{4r+4}q 
\end{eqnarray}
where $q=e^{2\pi i \tau}$.

%% file: conclusion.tex
\setcounter{section}{0}
\setcounter{subsection}{0}
\setcounter{equation}{0}

%
%

The recursion relations discovered \cite{Recur}\cite{CalegroRecur} improve
considerably the ability to evaluate explicitly the non-perturbative
instanton corrections to $\N=2$ super Yang-Mills theories.   One may
speculate as to the existence of similar recursion relations for the
strong coupling problem \cite{Lerche}\cite{dhokerstrong} in the absence
of an explicit S-duality.  Other possible directions would be in speculating on
the existence of simple recursion relations in the mirror symmetry
problems of calculating worldsheet instanton corrections to string theory 
\cite{Candelas}.

A deeper question is that of the existence of the recursion relations in the 
first place. One approach is that the instanton corrections to the 
prepotential of $\N=2$ super Yang-Mills is related to a topological field 
theory version of the same theory where the 
the Lorentz group and ${\cal R}$-symmetry are twisted into one
another \cite{Fucito1}\cite{Fucito2}.  Another question is that of
the existence of the role of integrable systems in the spectral curves, and
how it relates to topological field theory and D-branes \cite{dbrane}.

One is left to wonder if this "simplicity" that appears in Seiberg-Witten
theory is just the tip of the iceberg for even more exact calculations
in nonperturbative string theories.

%% file: bib.tex
%
%